# An insight into acupoints and meridians in human body based on interstitial fluid circulation

Hongyi Li [1*] Yajun Yin [2] Jun Hu [3] Hua Li [4] Fang Wang [1] Fusui Ji [1*] Chao Ma [5, 6 *]

[1] Cardiology Department, Beijing Hospital, National Center of Gerontology, Beijing, China

[2] Department of Engineering Mechanics, Tsinghua University, Beijing, China

[3] Key Lab of Interfacial Physics and Technology, Shanghai Institute of Applied Physics, Chinese Academy of Sciences, Shanghai, China; Shanghai Synchrotron Radiation Facility, Shanghai Advanced Research Institute, Chinese Academy of Sciences, Shanghai, China

[4] Key Lab of Intelligent Information Processing, Institute of Computing Technology and University of Chinese Academy of Sciences, Beijing, China

[5] Institute of Basic Medical Sciences, Chinese Academy of Medical Sciences, Department of Human Anatomy, Histology and Embryology, School of Basic Medicine, Peking Union Medical College, Beijing, China

[6] Chinese Institute for Brain Research, Beijing, China.

**\* Corresponding author**

Dr. Hongyi Li and Fusui Ji, Cardiology Department, Research Center for Interstitial Fluid Circulation & Degenerative Diseases and Aging, Beijing Hospital, No.1 Dongdandahua Road, Beijing 100730, China. Email: leehongyi@bjhmoh.cn (HyL) and jifusui0367@bjhmoh.cn (FsJ)

Dr. Chao Ma, Institute of Basic Medical Sciences, Chinese Academy of Medical Sciences, Department of Human Anatomy, Histology and Embryology, School of Basic Medicine, Peking Union Medical College, No. 5 Dongdansantiao, Beijing 100005, China. Phone: 010-69156469, Email: machao@ibms.cams.cn

**Abstract**

The atlas of human acupoints and meridians has been utilized in clinical practice for almost a millennium although the anatomical structures and functions remain to be clarified. It has recently been reported that a long-distance interstitial fluid (ISF) circulatory pathway may originate from the acupoints in the extremities. As observed in living human subjects, cadavers and animals using magnetic resonance imaging and fluorescent tracers, the ISF flow pathways include at least 4 types of anatomical structures: the cutaneous-, perivenous-, periarterial-, and neural-pathways. Unlike the blood or lymphatic vessels, these ISF flow pathways are composed of highly ordered and topologically connected interstitial fibrous connective tissues that may work as guiderails for the ISF to flow actively over long distance under certain driving forces. Our experimental results demonstrated that most acupoints in the extremity endings connect with one or more ISF flow pathways and comprise a complex network of acupoint-ISF-pathways. We also found that this acupoint-ISF-pathway network can connect to visceral organs or tissues such as the pericardium and epicardium, even though the topographical geometry in human extremities does not totally match the meridian lines on the atlas that is currently used in traditional Chinese medicine. Based on our experimental data, the following working hypotheses are proposed: 1, there are one or more ISF flow pathways, including at least one cutaneous pathway, originated from an acupoint on the body surface. 2, the acupoints on the body surface specifically connect with certain visceral organs or tissues via ISF flow. And 3, the acupoint-originated ISF pathways constitute a complex connective network and can modulate the ISF and bio-signals in the microenvironments around cells in certain visceral organs or tissues from body surfaces. A comprehensive atlas will be constructed to systemically reveal the detailed anatomical structures of the acupoints-originated ISF circulation. Such an atlas may shed light on the mysteries shrouding the visceral correlations of acupoints and meridians, and inaugurate a new frontier for innovative medical applications.

**Introduction**

As one of the human anatomical accounts recorded by ancient Chinese in the year of 1027, the Tian Sheng Bronze Statue showing acupoints and meridians atlas introduced a clinical guideline for medical practitioners in the Traditional Chinese Medicine (TCM) **[1, 2]**. During a long history, it has evolved to the currently used acupoints and meridians atlas. To date, the acupoints on the atlas represent defined areas on the body surface relative to certain landmarks **[2]**. Each meridian line of the 12 main meridians is usually a virtual line that connects the adjacent acupoints belonging to the same group depicting a direct connection from acupoints in the extremity ending to diverse associated visceral organs or tissues **[1, 2]**. Anatomically, the meridian lines are known to distinct from the cardiovascular or the nervous system. Functionally, meridians could transport "Qi and blood" to nourish the interior organs and extremities. However, in modern medical science, the anatomical structures and physiological functions of acupoints and meridian lines are not yet totally clarified.

For several decades, a number of research approaches have been adopted to identify the anatomical and histological structures for acupoints and meridian lines. As neither the blood vessels nor the nerves can match the meridian lines, scientists have been trying to find other structures. However, so far no one is able to find a conduit-like structure connecting the acupoints and extending from an extremity ending into different visceral organs **[3-4]**. Instead, some anatomical structures were found to correspond to acupoints, such as the peripheral nerve endings, connective tissue facial planes, and collagen fiber bands, etc. **[5-7]**. Although multiple studies have been published to support the acupuncture's efficacy or adjunct efficacy in various diseases **[8-10]**, the specificity of acupoints and their connections with the corresponding visceral organs have yet to be identified.

In TCM, an acupoint is considered as a gateway for substances or bio-signals to enter or escape the meridian **[1, 2, 11]**. Theoretically, by tracking the substances or bio-signals from an acupoint might be a way to identify the anatomical structures of meridians. Therefore, efforts were made to visualize the meridians by injecting appropriate tracers into the acupoint.

In this paper, we review previously published studies on visualizing the meridians, and summarize our recent discoveries of a long-distance interstitial fluid (ISF) circulatory pathway that may originate from the acupoints in the extremities. Based on our preliminary experimental data, working hypotheses are proposed at the end of this review, together with a proposal to construct a comprehensive atlas for the detailed anatomical structures of the acupoints-originated ISF circulation.

**Early explorations of fluid flow from an acupoint**

In the early 1960s, by hypodermic injection of a colored dye into the acupoints of animals, it was claimed that a conduit-like structure named "Bonghan ducts" could be observed for fluid flow from acupoints **[12]**. However, the observations cannot be reproduced. Since the 1990s, these structures have been re-investigated by the injection of Trypan blue or Alcian blue and was renamed as "primo-vessels" or "primo vascular system" **[13, 14]**. Despite of controversy, the "primo-vessels" basically represent a new kind of micro-conduit lined with endothelial cells to support fluid flow and contain multiple channels surrounded by loose collagenous matrices **[15, 16]**, and were found to distribute within or outside the blood and lymphatic vessels, on the surface of organs, in the brain ventricles and central canals, and in the skin, etc. **[17]**. However, the detailed anatomical and histological structures of the "primo-vessels", and the relationship with human acupoints have yet to be clarified.

In the 1950s-1990s, radionuclide imaging techniques have been used to visualize a long-distance migration channels by the hypodermic injection of an isotopic tracer into an acupoint in the hands or feet of humans **[11, 18-20]**. Meng et al. investigated the migration channels of technetium-99$^m$ from the acupoints of 12 main meridian channels and found that the isotopic tracer migrated over a long distance in either the upper or lower limbs **[21-23]**. The radiotracer migration from the acupoint of Neiguan were distinct from those from the non-acupuncture and non-meridian point **[24]**. The trajectories of the radioactive tracers from acupoints were also different from those of intravenously injected isotopes and lymphotropic isotopes **[19, 23, 25, 26]**. However, the anatomical distributions of the isotope migration channels in humans did not match precisely with those of the 12 main meridians on the atlas. The researchers explained that the discrepancies were due to individual differences among human subjects and the acupoints as well as meridians atlas is a schematic diagram **[11, 18, 19, 21-24, 27]**. Lazorthes showed that the radiotracer from the acupoint should enter the venous system via the lympho-venous anastomoses and therefore denied that the radiotracer migration proves the existence of the meridian lines **[28]**.

Further studies compared the isotope channels from an acupoint with the blood and lymphatic vessels in animals. Kovacs et al. used a canine model to demonstrate that the longitudinal progression of an isotope-labeled tracer was not via the blood nor lymphatic vessels by comparison with the migration trajectories of a technetium-99$^m$ tracer injected intravenously and a lymphotropic rhenium sulfide injected hypodermically **[25]**. The longitudinal migration of the isotopic tracer was prevented by the introduction of a deep cutaneous incision, and could be restored when the incision plane was filled with a transonic or silicone gel **[26]**. These findings suggested that the isotopic migration channel from an acupoint should not be a conduit-like vessel. Other studies explored the characteristics and mechanisms of the isotopic tracer from acupoints in animals, such as the low electrical resistance **[29]**, hydraulic resistance along meridians **[30]**, perivascular space **[31]**, higher partial oxygen pressure **[32]**, accumulated calcium ion **[33]**, neurotransmitter transmission **[34]**, etc. Although these studies strongly indicated that the radiotracer from an acupoint might flow via an extravascular pathway, the anatomical and histological structures of the isotope migration pathway could not be clearly identified in human subjects. Due to the poor spatial resolution (approx. 1cm) of scintigraphic images, it was also very difficult to clarify the relationship between the isotope migration pathways and the meridians.

Despite of numerous controversy and inconsistent results, the above studies have provided us valuable clues and preliminary observations for the fluid flow originated from the acupoints in humans or animals. To further investigate these phenomena, imaging techniques with higher spatial and temporal resolution are required. Owing to the significant progress of medical and molecular imaging during the past decades, we have adopted a number of advanced techniques including magnetic resonance imaging (MRI) and fluorescent imaging, and made a series of promising discoveries of ISF flow in living human subjects and cadavers **[35-38]**.

**Visualization of ISF flow pathways originated from the acupoints**

MRI has a higher spatial resolution than radionuclide imaging. By hypodermic injection of a paramagnetic tracer (Gd-DTPA) into an acupoint of the hands or feet in healthy volunteers and observed by MRI, the fluid flow originated from the acupoint could be clearly visualized as long-distance imaging trajectories with several smooth or non-smooth pathways in the extremities **[35, 36]**. The smooth pathways showed a continuous trajectory, whereas the non-smooth pathways showed a discontinuous trajectory **[36]**. If the injection point was not an acupoint but in the vicinity of a vein, only the smooth pathways were observed. Therefore, the smooth pathways appeared to be the vein, although it was difficult to determine that the exact position of the paramagnetic tracer was either inside or outside the blood vessels due to the limitations of MRI technique used. Further analysis of the imaging data showed that the non-smooth pathways from acupoints were in the subcutaneous tissues of the hands, forearms, feet, and lower legs and did not match any blood vessel [36]. Moreover, the lymphatic vessels from an acupoint were visualized by hypodermic injection of a lymphatic contrast agent, iodized oil into the acupoints in the right and left thumb [36]. By comparison, neither the smooth nor the non-smooth pathways from the same acupoints identified by MRI matched the lymphatic vessels. In accord with the previous findings in the radionuclide imaging, the paramagnetic tracer enhanced pathways from the acupoints in the MRI did not fully overlap with the main meridian lines on the atlas.

To identify the anatomical and histological structures of the ISF pathways from an acupoint in physiological conditions, fluorescent imaging was adopted in an *ex vivo* human leg sample under a very special circumstance [37]. In a patient with severe foot gangrene and preparing for lower leg amputation, a fluorescent tracer was injected hypodermically into the Kunlun acupoint at the ankle before the amputation, following a research protocol approved by the hospital ethics committee and the written informed consent. After the amputation, the lower leg was dissected layer by layer. A number of long-distance pathways were revealed by the fluorescent tracer, which originated from the Kunlun acupoint at the ankle and extended to the end of the amputated leg. Based on the histological analysis under fluorescent microscopy, the following 4 types of pathways were identified (see Figure 1A for examples and schematic drawing): 1, a cutaneous pathway in the dermis and hypodermis tissues; 2, a perivenous pathway along the venous adventitia (including the adventitia); 3, a periarterial pathway along the arterial adventitia (including the adventitia); and 4, a neural pathway that is composed of the endoneurium, perineurium, and epineurium of a peripheral nerve. Compared with the MRI results, the cutaneous pathways were probably the non-smooth pathways, and the perivascular pathways were probably the smooth pathways [36]. Regardless of the types, all the above 4 pathways originated from the acupoint were composed of fibrous connective tissues rather than endothelial cells. Taken together, these data provide confirmative evidence that the acupoint-originated long-distance fluid flow pathways are distinct from the blood or lymph flow and should be classified as a specific category of ISF pathway.

The ISF pathway from an acupoint was also investigated under non-physiological conditions in three amputated lower legs in human patients [37]. The fluorescent tracer was injected hypodermically into the same acupoint of Kunlun after the amputation. To visualize the long-distance ISF pathways, mechanical compressions were performed repeatedly using a sphygmomanometer cuff applied at the amputated end of the leg to achieve a systolic pressure of 50–60 mmHg and a compression-relaxation frequency of 18–20 times/min. After 90 mins of manipulations, the same 4 types of the fluorescently labelled ISF pathways were also observed. Without the extra mechanical compressions, the fluorescent tracer only diffused around the injection site and did not flow from ankle to the amputated end of the leg. Confocal laser microscopy showed that all the ISF pathways, regardless of physiological or non-physiological conditions, contained abundant fluorescently stained micron-sized fibers that were assembled longitudinally in the direction of the ISF pathways [37]. The above results not only further confirm the presence of acupoint-originated ISF flow pathways with at least 4 types of anatomical distributions in human, but also indicate that the ISF pathways are composed of highly ordered and topologically connected framework in interstitial fibrous connective tissues.

In order to detect the systemic distributions of the ISF flow pathways in visceral organs or tissues, a novel approach of extra mechanical driving forces was applied to human cadavers [38]. A mechanical automatic chest compressor was used on human cadavers to simulate the heart beating. Repeated chest compressions were performed for 2.5 hours after hypodermic injection of the fluorescent tracer into the Shaoshang acupoint in the first knuckle of the thumb. Fluorescent imaging revealed a long-distance ISF flow pathway from the right thumb to right atrium near the chest wall. In contrast with the 4 types of ISF flow pathways observed under physiological conditions, only 2 types of pathways were identified in cadavers (see Figure 1B for schematic drawing): a cutaneous pathway and a perivenous pathway. The cutaneous pathways from the thumb were found in the dermic, hypodermic, and fascial tissues of the hand and lower forearm, but not in the skin above the level of the cubital fossa. The perivenous pathways from the thumb were observed along the veins of the arm, axillary sheath, superior vena cava, and in the pericardium and epicardium tissues over the right atrium near the chest wall under the chest compressor. Histological analysis showed that the cutaneous and perivenous pathways were also composed of longitudinally oriented fibrous connective tissues. Moreover, the intrinsic structures of the cutaneous pathways from the acupoint were different from those of the skin tissues outside the transport pathways. Micro-CT imaging showed that the interlobular septum of the hypodermic adipose tissues is longitudinally assembled and oriented toward the transport along the cutaneous pathways. In contrast, the hypodermic interlobular septum of the skin tissues outside the pathways was irregular [38]. The above results showed that an acupoint on the hand anatomically connect with the pericardium and epicardium via a pathway that is composed of longitudinally oriented fibrous connective tissue. These findings are in accordance with the results from a similar experiment in the live rabbits **[39]**, and challenge the basic physiological concept that ISF is normally entrapped within the gel-like interstitial connective tissues and only occasionally flows freely as small rivulets and vesicles for a short distance in the interstitium. It could also be speculated that the mechanical beatings of the heart might be one of the factors that drive the flow of ISF systematically.

**Potential mechanisms underlying the long-distance ISF flow**

In conventional physiology, the explorations on ISF flow through interstitial connective tissues have started since 1896, when the Starling equation was formulated to described fluid exchanges across the capillary wall **[40-42]**. Using porous matrix theory and the measurements of hydraulic conductivity of various tissues based on a method designed by Guyton and Levick, it was found that ISF diffuses mainly for short distances through the tissue gel between capillaries and adjacent cells **[43-46]**. The interstitial permeability is mainly determined by the concentrations of both glycosaminoglycans and collagen in a wide variety of tissues **[47]**. If the ISF were to flow for long-distance, ISF must have been driven not only by the interstitial pressure changes but a continuously distributed interstitial pressure gradient that acts on consecutively adjacent connective tissues. However, techniques for direct measurement of ISF pressure under physiological conditions are usually invasive and limited in their capacity to determine the instantaneous interstitial pressure at different locations in the interstitium. Even if the interstitial pressure gradients would be established, it would be almost impossible for ISF flow to form a longitudinal pathway as observed in the human or animal experiments. The intrinsic structures and boundaries of interstitial connective tissues for a long-distance ISF flow must be clarified. Therefore, the use of imaging tracers is an alternative for direct observations of the ISF flow *in vivo*.

By observing the distribution of ferrocyanide ions after leaking from blood vessels, a water-rich region in the interstitial connective tissues was proposed and named the "interstitial tissue channel" by Hauck and Casley-Smith **[48-50]**. These tissue channels were proposed to act as a converging drainage pathway to supply or drain fluid in a unit of interstitial connective tissues. Although some scientists suggested that tissue channels may play a role in delivering ISF flow **[51]**, the spatial structures of both the tissue channel and its surroundings in the gel-like interstitial matrix for a long-distance ISF flow have not been clarified until now **[48]**.

Our experimental findings on human cadavers, the amputated lower legs and animals have disclosed a unique kinetic and dynamic pattern of ISF flow **[36-39]**. The reciprocating movements of a driving center, such as the heartbeat, are of vital importance to the flow of the ISF through these long-distance and spatially oriented fibrous pathways **[37-39]**. Although the dynamic mechanisms of ISF flow are not yet fully understood, it has been demonstrated that the ISF is "pulled" from the long-distance fibrous pathways into the driving center. The unique kinetic and dynamic pattern provided the following conclusions that may clarify the long-distance transport space for ISF flow through the interstitial connective tissues **[52, 53]**.

1. All the ISF flow pathways, including the cutaneous pathway, the perivenous pathway, the periarterial pathway and the neural pathway, are composed of fibrous connective tissues but not conduits formed by endothelial walls.
2. At microscopic level, the intrinsic framework of the ISF flow pathways comprised abundant micron-sized fibers that are assembled longitudinally toward the direction of transport.
3. The fluorescently stained fibers observed under the confocal laser microscopy are the results of the fluorescein mixed in the ISF flow through the long-distance non-vascular pathways.
4. Given that neither the fibers nor the gel substances in the fibrous connective tissues are flowable, there should be a long-distance transport space for fluid flow throughout the fibrous pathways.

From the viewpoint of interface science, an interfacial space would exist physically between two phases of matters **[54, 55]**. In the interstitial matrix, an interfacial zone would form physically between a solid phase (a fiber, a cell, a bundle of fibers, a group of cells, or a layer of fascia, etc.) and a liquid phase (the gel or liquid substance). The pores in an interfacial zone are tiny, probably a nano- or even micron-sized space paved on a solid surface. The physically formed interfacial zone might be a transport space for ISF flow through the interstitial matrix. Upon the longitudinally assembled fibers from the thumb to the superficial tissues on the heart, a long-distance interfacial transport zone (ITZ) might be topographically connected. **[5, 51].** The long-distance flow of the ISF may be driven by the heartbeat or other reciprocating movements via the topographically connected ITZs and diffuses into the surrounding gel substances along these pathways, where the longitudinally assembled fibers might play a role as a guiderail for the fluid flow in the highly ordered fibrous connective tissues. Confocal imaging of both the fluorescently stained fibers and gel in the fibrous matrix have provided *in situ* imaging evidence of an ITZ along the fibers **[37, 38]**.

It should be noted that an ITZ exists in multiple shapes overlaid on diverse solid structures in addition to a group of fibers. More patterns of ISF flow along topologically connected and highly ordered solid structures need be explored. The detailed kinetic and dynamic mechanism of ISF flow via the longitudinally assembled ITZs (named as interfacial fluid flow) also need further studies, especially under physiological conditions. The mechanisms of ISF flow through the long-distance fibrous pathways are suggested to be an interfacial fluid flow **[52, 53]**.

In conventional physiology, the interstitial connective tissues are thought to contain a chaotic framework and act as the "glue" for cells to attach to the matrix. Our data suggested that the fibrous framework of interstitial connective tissues is highly ordered and topologically interconnected; Via the topologically connected ITZs along the solid surfaces of a group of fibers, the ISF can flow through interstitial connective tissue toward a driving center. Thus, an ISF circulatory system has been proposed **[52, 53]**. Exchanged with plasma across the capillaries, ISF is not fixed in *tissue gel* but would circulate systematically around the whole body.

**Relevance between the acupoint-originated ISF pathways and the meridians**

According to our experimental results in human subjects, the topographical geometry of acupoint-ISF-pathways in hands or feet is significantly different from the main meridian lines on the atlas. Although the sample size is small, it clearly shows that the ISF flow pathways do not fully match the main meridian lines. However, there are some relevance between the acupoint-originated ISF pathways and the known functions of meridian lines.

Firstly, the transport pattern of the cutaneous acupoint-ISF-pathways is unique due to their specific tissue boundary. Unlike blood and lymphatic vessels, a long-distance cutaneous acupoint-ISF-pathways is not a hermetically sealed conduit but an opened flowing and irrigating pathway. Such a cutaneous pathway may transport ISF from an original acupoint, and collect or spread ISF from or into the surrounding tissues or organs along the long-distance transport pathways, and deliver the substances or bio-signals via the ISF into certain visceral organs under the act of driving centers. Previous studies have revealed that the acupuncture signals include biophysical stimulations and biochemical reactions **[56, 57]**. When the mechanical forces of the needle were applied on the connective tissues during twisting **[58]**, numbers of biochemical and biophysical signals were activated, such as various neural and neuroactive components **[59]**, neuroendocrine regulation **[60]**, durotaxis of cells (by affecting the mechanical properties of extracellular matrix) **[61]**, mast cells degranulation **[62,63]**, etc. Mediated by the long-distance ISF flow, both biochemical and biophysical acupuncture signals can be conducted.

Secondly, the acupoint-ISF-pathways from some Jing-well acupoints seem to be correlated with some downstream acupoints of meridians, like Hegu and Sanyinjiao acupoint that were visualized in volunteers by MRI (Figure 1C-D for 3D-reconstructed MRI image and schematic drawing). The non-smooth ISF flow pathways originating from Shangyang (the Jing-Well acupoint of Large Intestine Meridian of Hand-Yangming) were found to converge into an intersection point or pass by the downstream acupoint of Hegu in dorsal hand (Figure 1C). Originated from the three upstream acupoints of Yinbai in Spleen Meridian of Foot-Taiyin, Dadun in Liver Meridian of Foot-Jueyin, and Taixi in Kidney Meridian of Foot-Shaoyin, it was found that three non-smooth ISF pathways pass by the downstream acupoint of Sanyinjiao in different depths but not converge into an intersection point at one plane under the acupoint of Sanyinjiao (Figure 1D).

Thirdly, understanding the physiological functions of meridian lines on atlas needs to disclose the anatomical distributions of different types of the acupoint-ISF-pathways from the limb extremities to the visceral organs. The coordination of the cutaneous, perivascular, neural, and fascial ISF pathways with the skin, vascular system, nervous system, and fascia system needs to be clarified, respectively.

**The networks of the acupoint-ISF-pathways**

The ISF flow pathways from a Jing-Well acupoint (locates on the tips of fingers and toes and the original acupoint of the main meridians) and some acupoints located in wrist or ankle have been displayed in a few human subjects by MRI or anatomy. Until now, all the acupoints that we studied have found to be connected with at least one of the four types ISF flow pathways, including a cutaneous pathway. Theoretically, all the ISF flow pathways originating from each acupoint on all the 12 main meridian channels could be visualized in human subjects and comprise a complex 3D network that may be described by directed connective graph (DCG).

To analyze the acupoint related complex networks, a "node" can be corresponded to an acupoint and the "lines" to the ISF flow pathways. The types of "node-lines" can be defined according to the anatomical positions, such as a cutaneous node-lines, a perivascular node-lines, a neural node-lines, a facial node-lines, etc. Consequently, the ISF pathways from the acupoints on one main meridian line can be labeled. For example, the node-lines originated from each acupoint on the Lung Meridian of Hand-Taiyin can be labeled as the Hand Lung Meridian network of node-lines. When most of the acupoints along each main meridian line of a total of 12 meridians are visualized, there would be 12 meridian networks of node-lines. Each meridian node-lines network comprises two parts: the limb part and the visceral part.

Based on the experimental data, the DCG can be used to align and map the complex network of the anatomical node-lines. The reflected projections of the node-lines on the skin can be visualized and constructed as a topographical atlas of the categorized node-lines networks in the whole body. In addition, the physiological functions of the anatomical node-lines that are associated with the nervous, vascular, and fascial systems can be labeled as well. Also, advances in medical imaging techniques can facilitate a more effective and comprehensive description of this network.

**Perspectives on meridians in the lights of acupoint-originated ISF pathways**

Based on our preliminary experimental data, the following working hypotheses are proposed:
1. There are one or more ISF flow pathways, including at least one cutaneous pathway, originated from an acupoint on the body surface.
2. The acupoints on the body surface specifically connect with certain visceral organs or tissues via ISF flow.
3. The acupoint-originated ISF pathways constitute a complex connective network and can modulate the ISF and bio-signals in the microenvironments around cells in certain visceral organs or tissues from body surfaces.

A comprehensive atlas will be constructed to systemically describe the detailed anatomical structures of the acupoints-originated ISF circulation. In TCM, meridians form a complex network of connections among acupoints and visceral organs. The 12 main meridian lines are interconnected and mutually coordinated to modulate the biophysical signals and biochemical substances, if any, by working together with cardiovascular system, nervous system, motor system including the fascia, and other known body systems **[39, 64]**. Rather than simple lines connecting the adjacent acupoints, meridians may imply complex anatomical and functional connections between the acupoints on the body surface to diverse associated visceral organs or tissues. Our discovery of acupoints-originated ISF flow might be essential to understand the substance or signal transmission as conducted by meridians. Unlike the previous studies conducted to identify a long-distance conduit for ISF flow, our experimental data demonstrated that the interstitial connective tissues comprise a highly ordered and topologically connected three-dimensional framework, which may work as a guiderail for the ISF to flow actively over long distance under certain driving forces. The ITZs on the longitudinally assembled solid surfaces of fibers or other solid structures are proposed to constitute a long-distance transport space in the network of interstitial connective tissues.

Basically, ISF is the major component of biophysical and biochemical microenvironments around cells, tissues, and organs. The acupoint-originated ISF flow pathways form a bridge from an acupoint to a visceral organ or tissue anatomically. It is reasonable to conjecture that an ISF circulatory network is not only a fluid transport system but also an actively dynamic communication system. By means of mechanisms that are yet to be defined, such as a regular heartbeat, irregular respiration, and skin evaporation, the ISF circulatory system can transport, communicate and regulate ISF and bio-signal exchanges among each part of the human body **[52]**. The detailed mechanisms underlying ISF circulation and the interactions with other body systems will need further investigation.

In summary, we discovered acupoint-originated long-distance ISF circulation via the highly ordered and topologically connected interstitial connective tissues. A comprehensive atlas was proposed to systemically describe the detailed anatomical structures of the acupoints-originated ISF circulation, namely the Human Interstitial Fluid Connectome Atlas (HIFCA) **[52]**. Such an atlas may shed light on the mysteries shrouding the visceral correlations of acupoints and meridians, and inaugurate a new frontier for innovative medical applications. In line with previously published studies, the acupoint may become an historical intersection of ancient medical knowledge and modern medical science where we could cherish ancestral treasures and make fruitful insights on the future of life science and medicine. Based further experimental data, it might be possible to visualize a new human topographical atlas originating from the acupoints on Tian Sheng Bronze Statue after one millennium since 1027.

**Acknowledgements**

This work was financially supported by Beijing Hospital Clinical Research 121 Project (121-2016002) and National Natural Science Foundation of China (82050004, 82050005, 12050001), and the CAMS Innovation Fund for Medical Sciences (CIFMS #2017-I2M-3-008).